# Structure of Antiphase boundaries in Ni-M-Ga: multiscale modelling

Jan Zemen,[1,2] František Máca,[1] Václav Drchal,[1] Martin Veis,[3] Oleg Heczko[1]

[1] Institute of Physics ASCR, Prague, Czech Republic
[2] Faculty of Electrical Engineering, Czech Technical University in Prague, Czech Republic
[3] Faculty of Mathematics and Physics, Charles University, Prague, Czech Republic

**Abstract**

Antiphase boundaries (APBs) are ubiquitous in ordered Heusler alloys and strongly influence magnetic coercivity in Ni-Mn-Ga, yet the link between their atomic-scale exchange interactions and micrometer-scale magnetic contrast measured by magnetic force microscopy (MFM) remains unclear. We combine density functional theory (DFT) and finite-element magnetostatics to bridge these scales in Ni-Mn-Ga. DFT calculations on supercells containing planar APBs show that the lowest-energy configuration comprises a pair of parallel APBs enclosing a nanoscale region - only three Mn–Ga atomic layers thick - whose magnetization is antiparallel to the surrounding matrix due to strong antiferromagnetic exchange across each APB (in contrast to ferromagnetic coupling in bulk martensite). According to our magnetostatic finite element model, this thin region with antiparallel magnetization generates the characteristic MFM contrast extending ~100 nm from the APB pair. When the APBs are further apart than ~ 50 nm, dipole–dipole penalties outweigh exchange gains, preventing formation of an extended antiparallel domain, in agreement with experimental evidence. These results identify APB pairs as the origin of the observed MFM contrast and offer an interpretation of the modest strengths of domain-wall pinning by APBs, informing the design of magnetic shape-memory alloys with tailored coercivity.

**Introduction**

Antiphase boundaries (APBs) are planar crystallographic defects intrinsic to ordered alloys that play an important role in tuning mechanical, thermal and magnetic properties. APBs have been shown to induce elastic softening in alloys such as Fe-Al-Ti with L2$_1$ structure [1] or Ni-Mn-Ga with austenite and premartensite structure [2]. APBs in melt spun Fe$_2$VAl may contribute to lowering of thermal conductivity of the thermoelectric alloy [3]. In Heusler alloys it can be formed thermally upon the ordering transition [4] or by mechanical deformation [5].

Thermal APBs in magnetic shape memory alloys (MSMAs) affect magnetic coercivity through magnetic domain wall (DW) pinning [6, 7, 8, 9], which was ascribed to antiferromagnetic coupling of local magnetic moments across APBs in Cu$_2$MnAl alloys as early as 1973 [4]. Such exchange phenomena are not unique to MSMAs. For example, a DFT study of Fe$_3$O$_4$ showed antiferromagnetic coupling of spins across the highly stable APB defects in the (110) planes [10]. On the other hand, DFT calculations of APBs in closely related half-metallic Co$_2$Fe(Al,Si) have shown that the preference for ferromagnetic ordering across the APB remain as in bulk, in contrast to magnetite or Cu$_2$MnAl mentioned above. Although, the strength of the FM interaction at APBs in Co$_2$Fe(Al,Si) is significantly decreased, which leads to reduction of the width of magnetic DW pinned on the APB, as confirmed by at atomistic spin simulations [11]. In contrast to magnetic Heusler alloys where APBs suppress the ferromagnetic order, in related alloys such as Fe-Al with B2-type (CsCl-type) ordered structure, it has been shown using electron holography that the nm-wide APBs enhance ferromagnetism [12, 13, 14].

Detailed experimental investigation of APB structure at the nanoscale is complicated as there is no transmission electron microscopy (TEM) contrast on APBs [15]. The only visualization is provided by magnetic contrast, which does not deliver sufficient level of detail. Lorentz force TEM (L-TEM) studies have suggested complex character of APBs formed between L2$_1$ ordered structural domains [9, 15, 16, 17]. Further L-TEM studies [9, 15, 18], as well as magnetic force microscopy (MFM) [19] of Ni$_2$MnGa have shown a clear difference between signals generated by a thermally induced APB and by a single magnetic DW (Bloch wall) in modulated martensite exhibiting uniaxial magnetocrystalline anisotropy.

A change of sign of the out-of-plane component of magnetic flux ($B_z$) measured across the interface is a signature of a APB in an in-plane magnetized sample, whereas a simple but much stronger peak of $B_z$ indicates a single DW. The observed magnetic DW pinning at APBs has important implications for macroscopic magnetic properties [7].

Moreover, a recent neutron powder diffraction (NPD) study of $Ni_2MnAl_{0.5}Ga_{0.5}$ and $Ni_2MnAl$ with partial $L2_1$ structure has confirmed that APBs in this material inherently induce magnetization reversal due to short-range antiferromagnetic exchange coupling [20]. However, it is known that the $L2_1$ order cannot be fully achieved in $Ni_2Mn(Ga,Al)$ in which large structurally disorder regions were observed [21].

These L-TEM, MFM and NPD observations leave an open question what is the link between the exchange interactions at the nanoscale and the observed magnetic contrast at the microscale, yet it plays a critical role in determining the performance of MSMAs. In particular, applications in magnetic sensing and actuation strongly depend on coercive field, which has been shown to grow significantly with APB density in Ni-Mn-Ga [7]. Advancing our understanding of the interplay between APBs and magnetism is a complex research subject, which requires multi-scale modeling approaches that bridge atomic-scale structures and micrometer-scale magnetic domains.

In this work we perform density functional theory (DFT) calculations of supercells accommodating one or two APBs in cubic austenitic $Ni_2MnGa$ and analyse total energies for different positions of the APBs. For the most favourable APB configuration we perform a magnetostatic simulation of magnetic flux generated in an air domain around a micrometer-scale magnetic object using finite element method (FEM). Finally, we compare the simulated results to magnetic contrast measured by MFM on $Ni_2MnGa$ samples with martensite structure [19].

**Experimental**

Antiphase boundaries were observed in single crystal Ni-Mn-Ga exhibiting 10M modulated martensite structure at room temperature. The 10M phase occurs after martensitic, i.e., difussionless, displacive transformation from the parent cubic $L2_1$ ordered phase called austenite at about 315 K. To keep structural compatibility upon transformation the martensite is twinned. The example of twinning causing regular surface morphology is shown in Fig. 1a visualized by atomic force microscopy. Apart from the regular twinning profile, there is no other surface features and the surface is flat. The modulated structure is often approximated by pseudotetragonal lattice with short crystal c-axis being easy direction of magnetization with magnetocrystalline anisotropy about $1.7 \times 10^5$ J/m$^3$. This relatively strong anisotropy secures full alignment of magnetization along easy axis and broad parallel magnetic domains [22].

The APBs were observed by MFM at room temperature. The double line magnetic contrast on APBs arises due to stray magnetic field above sample [19]. In order to visualize weak contrast arising from ABP, single magnetic domain was selected in two adjacent twin domains. An example of observed surface with APB is shown in Fig. 1b. The surface is approximately (100) plane of single crystal with easy magnetization direction in plane. Such arrangement provides the strongest contrast for APBs. Moreover, the directions of magnetization in twin domains are perpendicular to each other, which provides additional features for APB visualization. The highest magnetic contrast on APBs arises when the magnetization is perpendicular to the APB lines (intersection of APB manifold with sample surface) and APBs are not visible when the magnetization is parallel to the APB lines. Thus, it seems that the lines do not form closed regions, despite their formation during transformation from partially disordered B2' phase to $L2_1$ ordered phase by nucleation and growth [23]. However, this is only an illusion created by the MFM contrast formation. By close inspection it can be inferred that the ordered regions separated by APBs are closed but highly irregular. Moreover, from Fig. 1 it can be inferred that the twin boundary is not affected by presence of the APBs.

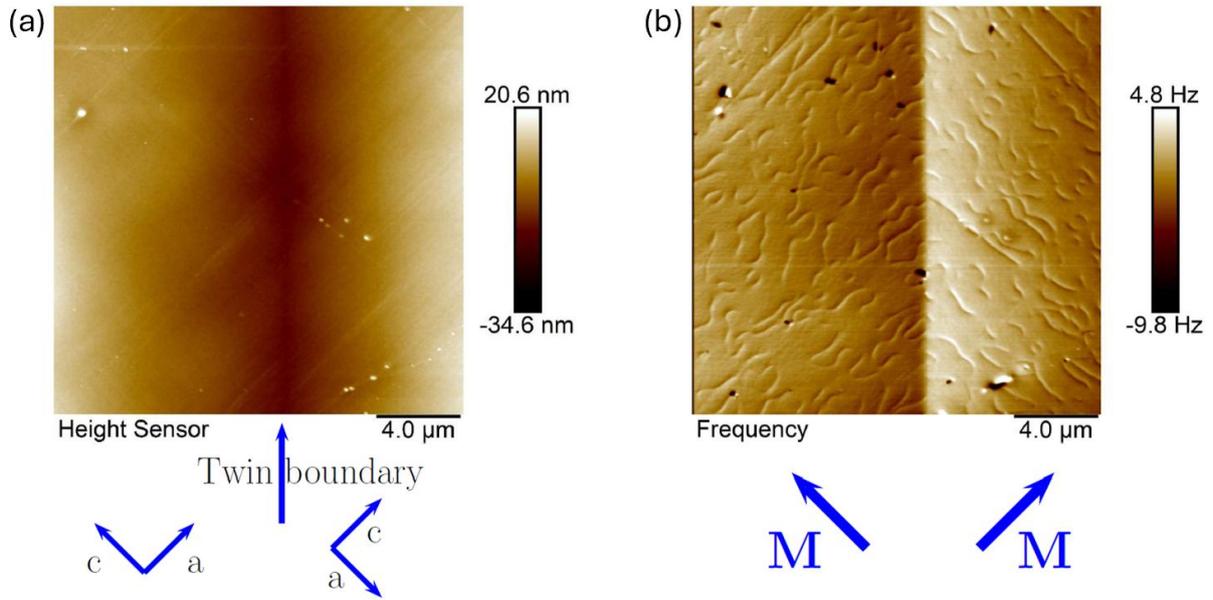

*Figure 1. Microscopy of the Ni-Mn-Ga sample surface parallel to (100) crytal plane with in-plane magnetization: (a) Atomic Force Microscopy showing surface morphology and (b) MFM providing magnetic contrast. The magnetic contrast arising due to vertical (perpendicular-to-surface) component of the magnetic induction, $B_z$. The positive and negative peak located at each APB is observed only when the plane of the APB is approximatelly perpendicular to the magnetization (indicated by blue arrows). Two orientations of magnetization in the sample are due to structural a/c twinning visualised in (a) as valley between two planes. The structure orientation abrubtly changes on the twin boundary.*

The signal is 100-1000 times smaller than the signal obtained for external field perpendicular to the surface. We note that magnetic DWs generate much larger magnetic contrast. The density of APBs is approximately 1 boundary per 400 nm. It is not clear what the angle between the APB plane and the sample surface is, but the low variation of the magnetic contrast across many APBs suggests that the contrast is proportional only to the angle between the magnetization and a tangent to the APB.

**Theory and modelling**

Our *ab initio* modeling is focused on the ferromagnetic Heusler alloy $Ni_2MnGa$ with cubic $L2_1$ structure [24]. We used supercells which are multiples of the unit cell along [110] crystal axis with one or two APBs included. The plane of each APB is perpendicular to the [110] axis of the original unit cell or the longer side of the supercell. An example supercell with 80 atoms and two APBs is shown in Figure 2. Such supercell is periodically repeated in all three directions without adding any vacuum gap, following the modeling approach of Lazarov et al. used in a DFT study of $Co_2Fe(Al,Si)$ [11]. The suppression of the degree of atomic order observed in the vicinity of APB, e.g., in $Ni_{50}Mn_{20}In_{30}$ by high-angle annular dark-field scanning transmission electron microscopy (HAADF-STEM) [25] is neglected in our modelling in order to require a reasonable amount of computational resources.

Taking the supercells analogous to Fig. 2 as input, we perform spin-polarized DFT simulations using the projector augmented wave (PAW) method as implemented in the Vienna Ab initio Simulation Package (VASP) [26], with generalized gradient approximation (GGA) of the exchange correlation functional as parameterized by Perdew-Burke-Ernzerhof [27]. We do not use Hubbard *U* to account for electronic correlations on Mn sites to avoid additional input parameters of our *ab initio* simulation. The valence configurations of Ni, Mn, and Ga are $3p^63d^94s^1$, $3p^63d^64s^1$, $3d^{10}4s^24p^1$, respectively. A reciprocal space mesh with $\Delta k = 0.05/Å$ is used to sample the Brillouin zone. Our plane wave energy cut-off energy is $E_{cut}$ = 600 eV. The total energy of the model systems calculated self consistently with convergence criterion, $\Delta E = 10^{-7}$ eV.

We start our simulations by relaxation of the atomic positions within the supercell with fixed volume and shape. This scalar-relativistic collinear spin-polarized calculation with residual force criterion, $\Delta F < 2$ meV/Å is repeated for supercells analogous to those in Fig. 2 with 28, 48, and 80 atoms. In case of the smallest supercell with 28 atoms and only one APB, we are able to carry out a subsequent simulations step with spin-orbit coupling in order to evaluate the magnetocrystalline anisotropy (MCA). In the case of supercells with 48 and 80 atoms, we include two APBs and explore the total energy as a function of separation of the adjacent APBs as shown in Fig. 2.

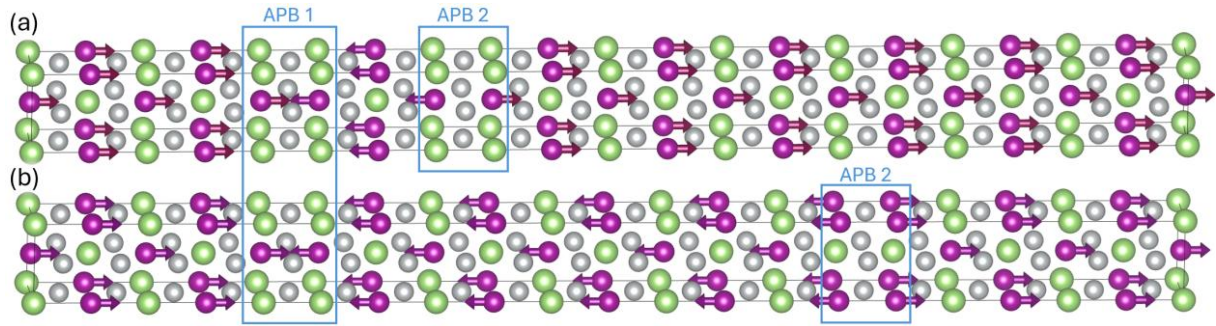

*Figure 2. Supercell of Ni$_2$MnGa with 80 atoms and two APBs (modelled as vertical planes); Mn atoms are in magenta, Ga atoms are in green and Ni in grey; All magnetic moments are horizontal (Mn moments are indicated by arrows): (a) separation of APBs by one Mn-Ga layer, i.e., the region with antiparallel magnetization is 3 Mn-Ga layers thick, (b) maximum separation of APBs, i.e., the region with antiparallel magnetizatio is 10 Mn-Ga layers thick.*

In order to compare our DFT results to the magnetic contrast obtained by MFM, we perform a magnetostatic simulation using the finite element method (FEM) as implemented in Comsol Multiphysics [28].

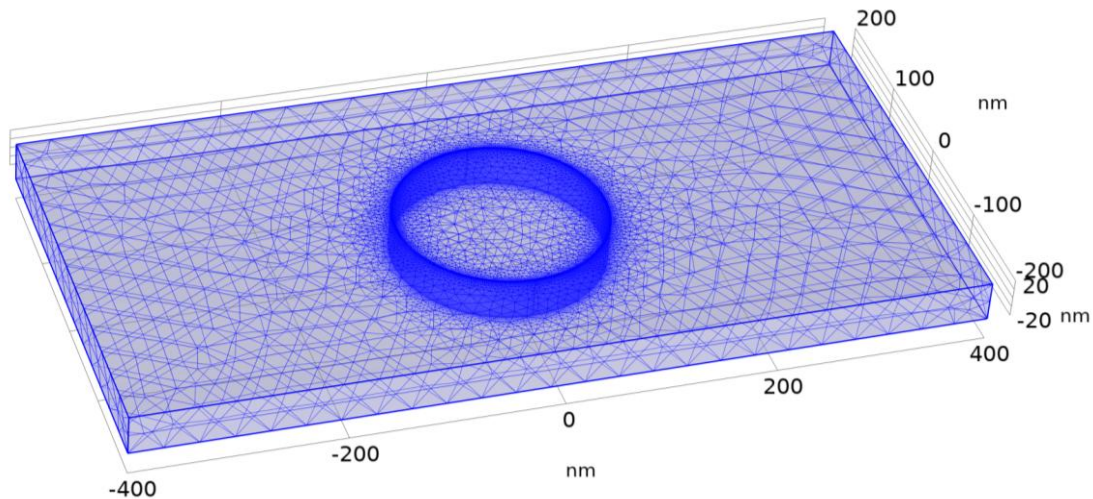

*Figure 3. Geometry and mesh of a Ni$_2$MnGa sample with a pair of APBs forming a ring. The mesh is much denser in the cylindrical domain as its width is only 1.74 nm, based on the DFT simulations. The film is homogenously magnetized along the x-axis (longer edge) with the exception of a narrow ring-shaped domain between the two APBs, which has zero magnetization. The film is surrounded by a large air domain with magnetic insulation boundary condition (not shown in the plot).*

Fig. 3 shows a ferromagnetic rectangular prism (Ni$_2$MnGa sample, 800 x 400 x 40 nm) with uniform magnetization, M = 755 kA/m (~0.95T) according to our DFT simulation, set parallel to the long side of the prism. There is a cylindrical gap in the material with radius of 100 nm and width of 1.74 nm, which is filled with vacuum (zero susceptibility, zero magnetization). This gap is used to model a pair of planar APBs identified as the most favorable ordering by our DFT simulations. The whole object is surrounded by a large vacuum domain (1600 x 800 x 600 nm) which is not shown in Fig. 4. The boundary of this domain is sufficiently far from our prism that we can use the magnetic insulation boundary condition. Our tetrahedral mesh is generated automatically taking into account the small size of the non-magnetic cylinder. It has over $10^6$ elements and we have checked the convergence of our results with respect to the density of the mesh. The default generalized minimum residual method is used to solve the system of linear equations.

**Results and Discussion**

We start by looking for the ground state of Ni$_2$MnGa with three different supercells (28, 48, and 80 atoms). Our spin-polarized DFT simulations identify the pristine L2$_1$ bulk phase (supercell without any APBs) as the most favorable lattice structure as expected. Systems with one or two APBs have higher total energies. The thermal APBs form during the order-disorder transition as the Ni-Mn-Ga alloy transforms from a partially disordered (B2 or B2') phase to the fully ordered L2$_1$ phase at 1070 °C [23]. The ordering concerns only Mn and Ga atoms as the Ni sublattice is formed upon solidification. The long-range ferromagnetic order appears at much lower temperatures around 100 °C. The predicted net magnetization of the bulk L2$_1$ structure is 15.868 $\mu_B$/c.c. (per conventional cell (c.c.) with a = 0.58 nm) which corresponds to M = 755 kA/m or 0.95 T as mentioned above, which is in good agreement with experimental values for Ni$_2$MnGa [29].

As the last check before studying supercells with non-uniform magnetization, we explore the effect of APBs on magnetocrystalline anisotropy (MCA), which could in principle explain the experimental dependence of magnetic coercivity on the number of APBs in a sample [6, 7, 8, 9, 19]. We compare the total energy of the smallest supercell (28 atoms and one APB) for parallel and perpendicular direction of the local magnetic moments with respect to the plane of the APB. Table 1 shows that the resulting MCA favours moments perpendicular to APB plane, i.e., parallel to the sample top surface in agreement with experiment. However, the MFM and L-TEM measured data do not provide any evidence of a preferential alignment of the APB manifolds with the sample top surface. At the same time, the simulated MCA of 0.013 meV/4 atoms (K$_u$ = 37.8 kJ/m$^3$) is rather small compared to uniaxial MCA in the 10M martensitic phase (K$_u$ = 170 kJ/m$^3$ [30]) so we must explore other atomic-scale mechanisms how the APBs can affect the coercivity.

| APB type | E$_{tot}$ [eV] | MCA [$\mu$eV/FU] |
|---|---|---|
| A$^{\rightarrow}$B$^{\rightarrow}$ | -174.826858 08 | 0.00 |
| A$^{\rightarrow}$B$^{\rightarrow}$<u>A$^{\rightarrow}$A$^{\rightarrow}$</u>B$^{\rightarrow}$A$^{\rightarrow}$B$^{\rightarrow}$ | -174.648237 34 | |
| A$^{\uparrow}$B$^{\uparrow}$<u>A$^{\uparrow}$A$^{\uparrow}$</u>B$^{\uparrow}$A$^{\uparrow}$B$^{\uparrow}$ | -174.648146 73 | 12.94 |
| A$^{\rightarrow}$B$^{\rightarrow}$A$^{\rightarrow}$<u>B$^{\rightarrow}$B$^{\rightarrow}$</u>A$^{\rightarrow}$B$^{\rightarrow}$ | -174.648237 88 | |
| A$^{\uparrow}$B$^{\uparrow}$A$^{\uparrow}$<u>B$^{\uparrow}$B$^{\uparrow}$</u>A$^{\uparrow}$B$^{\uparrow}$ | -174.648146 84 | 13.01 |

Table 1. Total energies calculated by DFT with spin-orbit coupling for relaxed supercells with 28 atoms and one APB. Atomic layers are indicated by letters A and B. The first row shows the total energy of a supercell without any APB, which is the ground state. Two letters A or B next to each other indicate the position of APB. Formation unit (FU) includes 4 atoms.

Experimental evidence suggests that coercivity changes are due to DW pinning at APBs, which have different magnetic character. It can be explained either by antiferromagnetic (AFM) exchange coupling across a single APB [4, 20] or by suppressed ferromagnetic (FM) exchange coupling close to the APB [11] or by AFM exchange inside a more complex APB region. Our MFM study of $Ni_2MnGa$ [19] has revealed that the magnetic contrast of a suspected APB and a 180° DW (gradual rotation of magnetization of Bloch or Neel type) are easily distinguishable. The field component perpendicular to sample surface, $B_z$, measured about 20 nm above the surface of an in-plane magnetized sample is mostly zero with a small negative and positive peak about 100 nm from the suspected APB, whereas the DW shows one strong peak in $B_z$. Therefore, we suggest that the magnetic contrast of the "thermal APB" type must be generated by a more complex structure than a single plane with magnetization reversal induced by AFM exchange coupling localized at an APB. The simplest magnetic structure that could explain the observed MFM signal is a pair of APBs close to each other and a narrow region between them with magnetization antiparallel to rest of the supercell. Such two planes with magnetization reversal close to each other would easily annihilate if they were not stabilised by APBs. Therefore, we simulate supercells with two APBs and a region between them magnetized antiparallel to the rest of the supercell, see Fig. 2. This structure would satisfy the expected AFM exchange coupling across each APB and does not require a DW located at the periodic boundary of the supercell or anywhere else in the defect-free $L2_1$ structure.

We perform spin-polarized DFT simulations of total energy for supercells with 48 and 80 atoms and two APBs as shown in Fig. 2. For each size, we vary the separation between the two APBs as shown in Fig. 2. The magnetic structure is initialized in a collinear state with moments reversed at each APB. During the DFT-relaxation of the atomic and magnetic structure, we do not observe any change in direction of the moments and only insignificant changes in the size of the local moments. Two examples of the magnetic moments on each atom are given in Fig. S1 for two different separations of the APBs. Contrary to our expectation, the magnitude of the moments is not affected by proximity to the APB and to the region with opposite magnetization. All moments in the complex magnetic region between the APBs are antiparallel to moments in the rest of the samples.

The total energies of the relaxed structures with two APBs are presented in Fig. 4. We compare a uniformly magnetized structure (despite the two APBs) and a structure with magnetization reversal at each APB. The first key finding is that the total energy of the uniformly magnetized system is approximately 10-20 $kJ/m^3$ higher than in the system with the complex magnetic region between ATPs (assuming approximately 1 APB per 400 nm according to MFM data in Fig. 1). In other words, our DFT simulations confirm the expectation of an AFM exchange coupling between Mn atoms at opposite sides of an APB, where their distance is shorter than in pristine $L2_1$ structure [4, 20]. This result also implies that the DW pinning at APBs observed experimentally [6, 8, 9, 19] is most likely due to the local AFM exchange rather than due to suppressed FM exchange or exceptional MCA around the APB. However, we caution that our DFT simulations predict an abrupt reversal of magnetization at each APB, there is no gradual rotation of magnetization typical for Neel or Bloch DWs in ferromagnetic materials. The formation of such regions with opposite magnetization is not driven by minimization of the classical magnetostatic energy, but by minimization of exchange energy instead. The remaining question is the origin of the magnetic contrast measured by MFM at suspected APBs [19], which cannot be explained by a single APB with magnetization reversal.

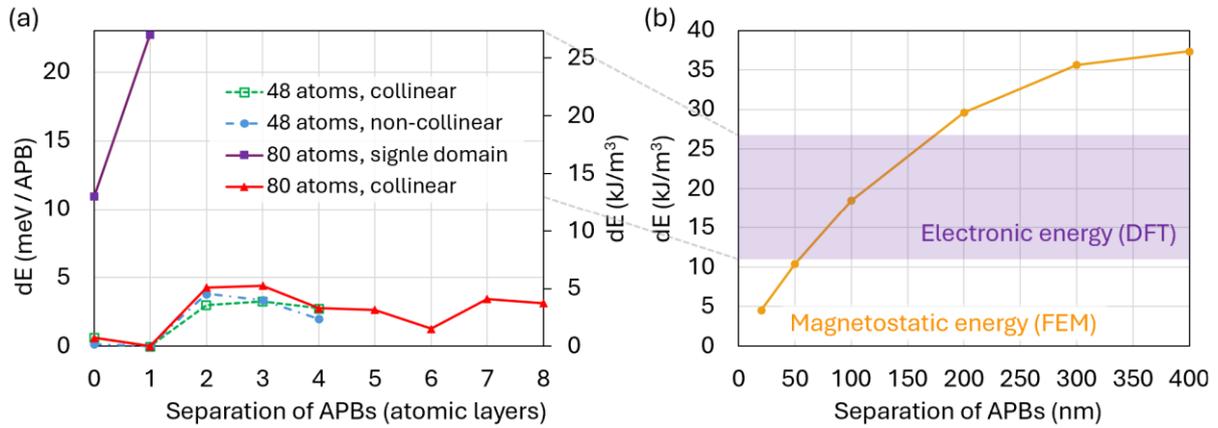

*Figure 4. Comparison of energy differences obtained by DFT and FEM; (a) electronic energy (DFT) – comparison of supercells with different separations of two APB two magnetic arrangements: supercell with homogenous magnetization and a supercells with magnetization reversal at each APB; The homogenous magnetic structure is approx. 10-20 kJ/m³ higher in energy than the structure with complex magnetic region, assuming 1 APB per 400 nm. The maximum separation of APBs is 4(8) Mn layers due to limited size of our supercell which is 48(80) atoms. (b) Magnetostatic energy (FEM) for different separations of APBs; The reference energy is the magnetostatic energy of uniformly magnetized sample.*

The second key finding presented in Fig. 4a is the optimal separation of APBs – the lowest energy is predicted for the structure with only one Mn-containing layer between the two APBs. This could be interpreted as a preference of ferromagnetic $Ni_2MnGa$ to form two APBs next to each other. However, the thermal APBs are formed by first order transformation at high temperature [23] and at much lower temperatures, when the magnetic order develops, any rearrangement of APBs is limited.

In order to interpret the MFM measurements, it is important to note that the total energy in Fig. 4a does not keep growing with increasing the size of the antiparallel magnetic region between two APBs. This is expected due to the exchange-driven formation of the antiparallel region (short-range interaction) and efficient electronic screening of structural defects in a metal. Now we will discuss separately two scenarios in order to propose hypothetical nanoscale structures underpinning the microscale magnetic contrast measured by MFM.

(i) **Complex ABP region:** Our DFT model can approximate this scenario only by two planar APBs with a thin antiparallel magnetic region squeezed between them, but the strong AFM exchange coupling across an APB responsible the antiparallel region is expected to be relevant in more complex APB configurations beyond our model. Despite its simplicity, our model can explain the observed MFM contrast. We note that our DFT simulations summarised in Fig. 4a do not show any perpendicular-to-surface component of magnetization. The size of the atomic magnetic moments does not change noticeably at the APBs as indicated in Fig. S1. We ran the relaxation in non-collinear regime, but the moments remained collinear. We even tried to model the surface by a slab and a vacuum gap in a supercell shown in Fig. S2 using Quantum ATK [31] but we could not find any non-collinearity. However, the MFM measurement is sensitive only to the vertical component of magnetic flux, $B_z$. Therefore, we complement our DFT simulations with a continuum model of the long-range dipole-dipole interactions (magnetostatics) for a realistic geometry of a sample surrounded by an air domain shown in Fig. 3. This FEM simulation allows us to predict the distribution of $B_z$ 20 nm above the surface of the sample, where it is measured by MFM presented in Fig 1.

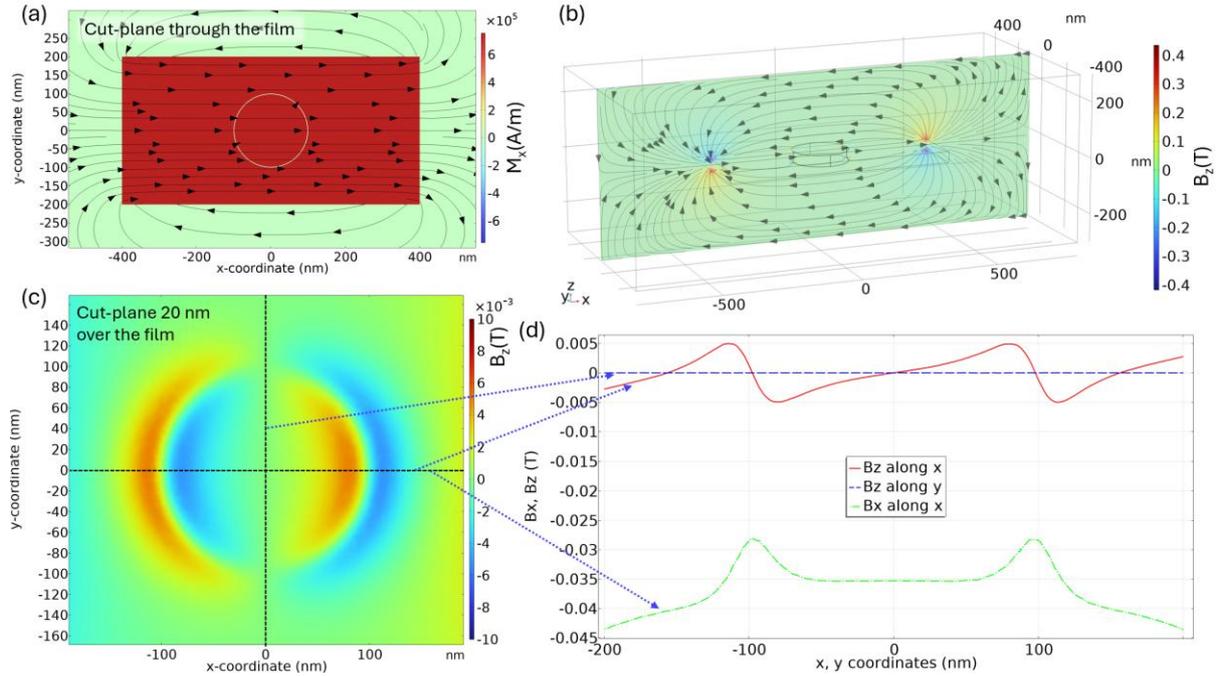

*Figure 5. Magnetostatic simulation of a sample surrounded by an air domain. The pair of APBs in close proximity is simulated by a ring-shaped domain with zero magnetization: (a) shows the top view of the sample with constant magnetization along the x-axis (in the plane of the top surface) and the flux lines in a cut-plane showing the corresponding magnetic induction **B**; (b) A vertical cut-plane through the sample and the surrounding air domain; (c) $B_z$ in a cut plane 20 nm above the sample top surface; (d) Cross section of panel (c) along the cut-lines indicated by black dashed lines.*

Fig. 5 presents our magnetostatic simulations of a macroscopic sample with a nano-sized APB region aiming to interpret the MFM data of Fig. 1. Fig. 5 (a) shows the top view of the sample with constant magnetization along the x-axis (in the plane of the top surface, which is the easy direction) and the flux lines in a cut-plane showing the corresponding magnetic induction, **B**. A vertical cut-plane through the sample and the surrounding air domain are shown in panel (b). The flux lines do not show a significant deviation from a pattern expected for a uniformly magnetized ferromagnetic rectangular prism. However, there is a ring defined within the sample as shown in panels (a) and (b). This ring has a radius of 100 nm and thickness of 1.74 nm, which describes the pair of APBs (wound into a ring shape) with the antiparallel magnetic region between them (identified by our DFT simulations). Here we model this complex APB region by a domain with prescribed zero magnetization, but it could be simulated by an even thinner domain with antiparallel magnetization. Fig. 5 (c) shows $B_z$ in a cut plane 20 nm above the sample top surface. We can see that despite the perfectly collinear magnetic structure with zero perpendicular-to-surface components, $M_z = 0$, the resulting magnetic induction has nonzero perpendicular component, $B_z$. Fig. 5 (d) then shows $B_z$ and $B_x$ in a cross section of panel (c) along the cut-lines indicated by black dashed lines. Strikingly, the spatial extent of the positive and negative peaks in $B_z$ is significantly larger than the thickness of the complex APB region in the sample underneath. Our cut-line along the y-axis of Fig. 5 (c), which is perpendicular to the magnetization, reveals that no magnetic contrast is observed for pairs of APBs that are perpendicular to the magnetization, which is in perfect agreement with the MFM data in Fig. 1. The size of the simulated $B_z$ peaks in Fig. 5 (d) is approximately 200 x smaller than the surface magnetization of the sample (measured in perpendicular alignment) – this ratio is in good agreement with the measured MFM data showing that APBs are not observable if easy magnetic axis and thus magnetization is perpendicular to the surface [19].

Moreover, if the two APBs were further apart than one Mn-layer, then the $B_z$ peaks would grow significantly as shown by simulations summarize in Fig. 6. In this case, the thickness of the ring grows from 1 to 50 conventional unit cells of the L2$_1$ structure and the region between APBs is modelled by magnetization fixed antiparallel to the rest of the sample, following the approach of DeGraef et al. [15]. The B$_z$ profile along cut-lines 20 nm above the sample surface (analogous to Fig. 5d) is shown in Fig. 7 for magnetization parallel, **M** = (M$_x$,0,0), and perpendicular, **M** = (0,0, M$_z$), to the sample surface, where |M$_x$| = |M$_z$| = 754.6 kA/m.

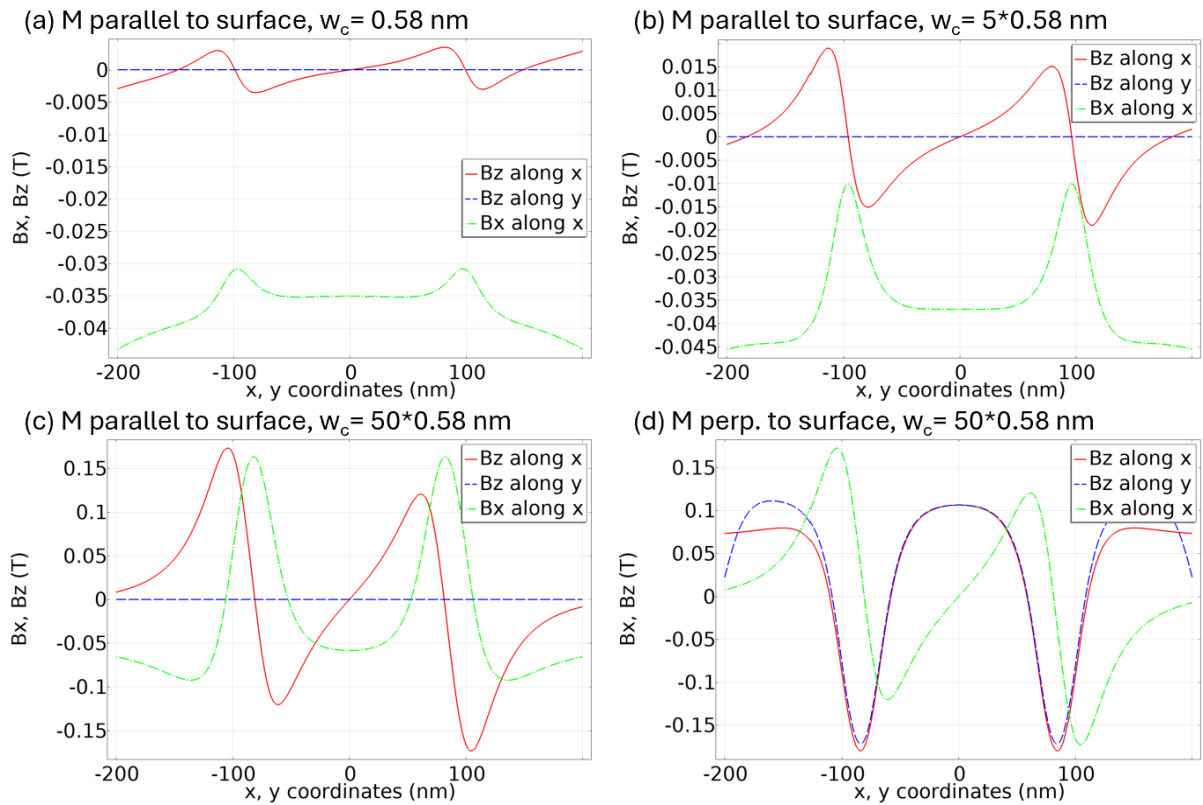

*Figure 6. Dependence of magnetic flux components on the width of the ring-shaped domain between two APBs. Here the region is modelled as magnetic with prescribed magnetization antiparallel to the rest of the sample. The size of the peaks increases significantly with growing width.*

So our first scenario offers an elegant interpretation of the MFM measured data based on the hypothesis that each line combining black and white contrast, originally suspected to be a single APB, is a pair of APBs very close to each other. However, our modeling is not capable of capturing more complicated APB manifolds so we interpret the MFM contrast more generally as a narrow APB region with dominant antiferromagnetic exchange interactions.

Nevertheless, our ambition is to interpret magnetic contrast due to thermal APBs formed during growth of the crystal. Under those conditions, it is possible that the short-range AFM exchange interaction would induce short-range ordering of atoms, e.g., drawing two existing APBs closer to each other. On the other hand, it is unlikely that it would completely prevent the existence of isolated APBs in the sample. Such APBs separated by more than one Mn layer from another APB cost more electronic energy according to Fig. 4a, but they would also generate much larger magnetic contrast according to panels *(b)* and *(c)* of Fig. 6. However, we do not see order of magnitude variations in the MFM magnetic contrast generated by different APBs in Fig. 1 or in related literature.

(ii)    **Isolated ABPs:** Continuum magnetostatic model of pairs of planar APBs with larger separations. In this scenario APBs are separated by two or more Mn-Ga layers. The predicted electronic energy is higher, but this is not relevant due to the thermal origin of these APBs. The magnetostatic simulations of Fig. 6 predict a much higher perpendicular component of magnetic flux, B$_z$, compared to scenario (i), which is not observed by MFM in any available samples. Therefore, we perform additional magnetostatic simulations of 10 isolated planar APBs in a 4x4 μm square sample shown in Fig. 7d, closely approximating the measured sample shown in Fig. 1 - the density is one APB per 400 nm. We respect the strong AFM exchange coupling across each APB so there are alternating stripe domains with $M_x>0$ and $M_x<0$ throughout the sample. The magnetization switches abruptly at each APB without forming a domain wall. We vary the distance between pairs of neighboring but relatively distant APBs and calculate the magnetic flux profile along a cut-line 20 nm above the sample, as shown in Fig. 7. Moreover, we evaluate the magnetostatic energy (taking the energy of uniformly magnetized sample as a reference), plotted in Fig. 4b. The magnetostatic energy grows significantly as the domains with antiparallel magnetization increase in width up to 400 nm. This classical energy contribution due to dipole-dipole interactions grows above the electronic energy predicted by DFT for the uniformly magnetized sample in the presence of APBs (again assuming 1 APB per 400 nm) as shown in Fig. 4.

Therefore, we predict that the antiparallel domain does not form between APBs separated by more than approximately 50 nm, because the exchange energy gain becomes fully compensated by the magnetostatic penalty. In other words, the isolated APBs are invisible in the magnetic contrast measured by MFM as they do not induce any inversion of magnetization direction. APBs closer than 50 nm to each other may induce an antiparallel magnetization but the resolution of the MFM may not be sufficient to distinguish such object from a pair of APBs of scenario (i). At the same time, we point out that a sequence of isolated APBs generates positive and negative peaks in $B_z$, as shown in Fig. 7 (a-c), but the peaks are very broad compared to scenario (i) – in contradiction to the measured magnetic contrast. This analysis is relevant for magnetization parallel to the sample surface due to magnetocrystalline and shape anisotropy. The energy with respect to the external field is not considered in our FEM simulation. In case of perpendicular magnetization, we expect the magnetic domain structure to be stabilized by the external field or strong magnetocrystalline anisotropy. The MFM signal is orders of magnitude larger for perpendicular magnetization and the contrast due to APBs is not visible.

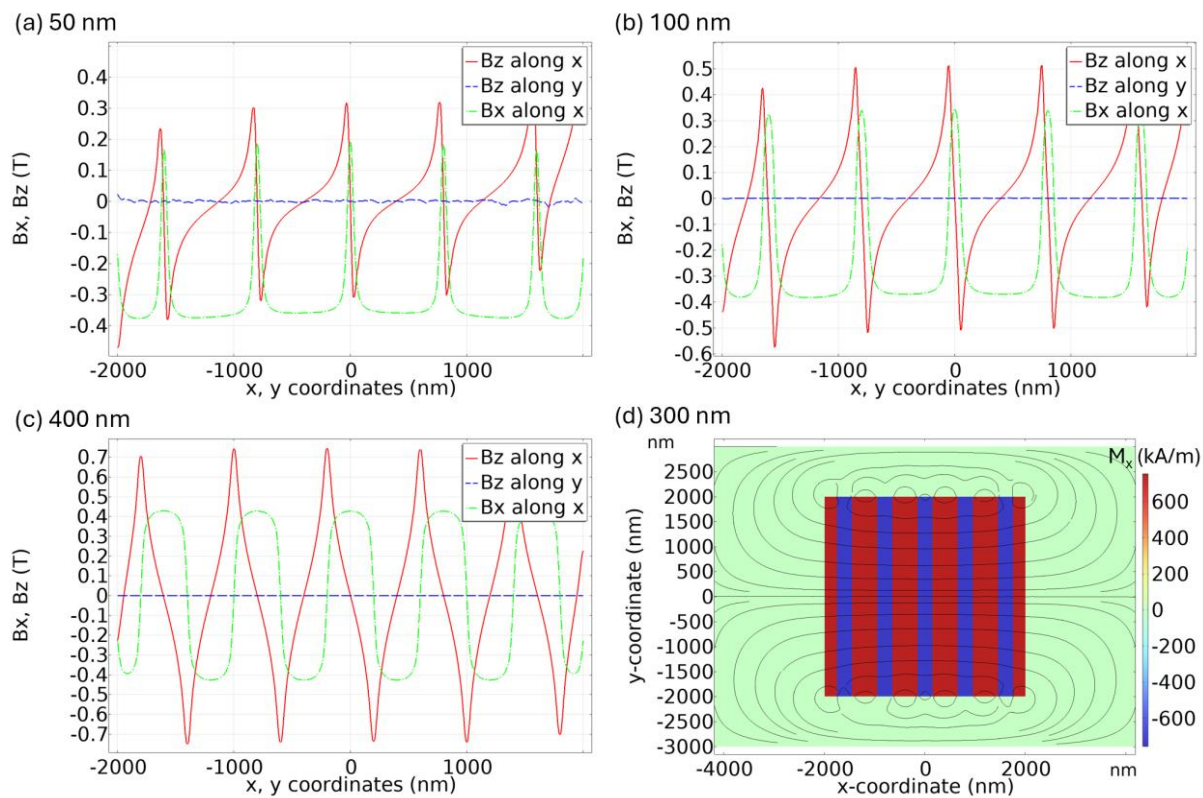

Figure 7. Magnetic flux profile of a square sample with 10 APBs. FEM simulation of magnetostatic energy with varying distance between adjacent pairs of APBs: (a-c) magnetic flux along cut-lines 20

*nm above the top surface; (d) Magnetization, $M_z$, along cut-plane through the sample. The corresponding predicted energy densities are plotted in Fig. 4b.*

**Conclusion**

Modelling of APBs in Ni-Mn-Ga on atomic scale and micro-scale was carried out to interpret magnetic contrast measured by MFM based on fundamental electronic interactions. DFT calculations of supercells accommodating one or two APBs in cubic austenite $Ni_2MnGa$ were performed. Our total energy analysis reveals that the lowest-energy APB configuration is formed by two parallel boundaries with a thin magnetic region between them with magnetization opposite to the rest of the supercell. The thickness of this region is only three Mn-Ga atomic layers. Despite being nanoscale, this region generates magnetic contrast with positive and negative peak in the perpendicular-to-surface magnetic induction, $B_z$, extending 100 nm around the pair of APBs. The magnetization of the sample is approximately 200 x larger than the maximal magnetic induction component $B_z$ predicted 20 nm above the sample surface. These characteristics are in quantitative agreement with our earlier MFM measurements [19] and with data in Fig. 1. Our DFT results are in line with the experimental evidence that APBs affect the coercivity via domain wall pinning. The underlying mechanism is the strong AFM exchange interaction between Mn atoms on opposite sides of an APB, while the exchange interactions between neighbouring Mn atoms in bulk martensite are ferromagnetic. However, due to much smaller thickness of the APB region compared to a Bloch domain wall, the pinning is not very effective.

The main result is the identification of curves with dark and light MFM contrast measured in Ni-Mn-Ga with pairs of APBs rather than individual APBs. The validity of our hypothesis is limited by the planar approximation of the complex APB structure. In general, the observed magnetization above the sample surface results from the balance of the long-range dipole-dipole interaction and short-range exchange interaction. When two APBs are close to each other, then the exchange interaction dominates and the narrow region between APBs magnetized antiparallel to the surrounding sample is formed and detected by MFM as a wavy line with dark and light contrast. However, when the APBs are far apart, then the dipole-dipole interaction penalty would exceed the gain in exchange energy so the larger antiparallel-magnetized region (domain) does not form. As a result, isolated APBs are not observed in magnetic contrast, which is an experimental fact. The hypothesis of APB pairs with magnetization reversal at each APB is in agreement with the seminal work by Lapworth and Jakubovics on Cu-Mn-Al in 1973. In our future work, we will attempt to explain the interaction of regular magnetic domain walls (not stabilised by AFM exchange coupling) and APBs which is an important aspect of DW pinning and has important implications for macroscopic magnetic properties of shape memory alloys.

**Supplementary/Appendix**

The size and sign of magnetic moments relaxed by VASP are shown in Fig. S1 for the supercell with 80 atoms and two separations between APBs. Each bar of the histogram describes one magnetic moment. The horizontal axis gives the position of the atom along the long axis of the supercell. The Mn moments are the largest, Ni moments are smaller and parallel to Mn moments. Ga moments are the smallest and antiparallel to Mn and Ni moments. Any atomic moments do not decrease in magnitude noticeably at the APB where the direction of the moment is reversed. Therefore, the local variations of the size of atomic moments cannot explain the measured pattern of magnetic induction.

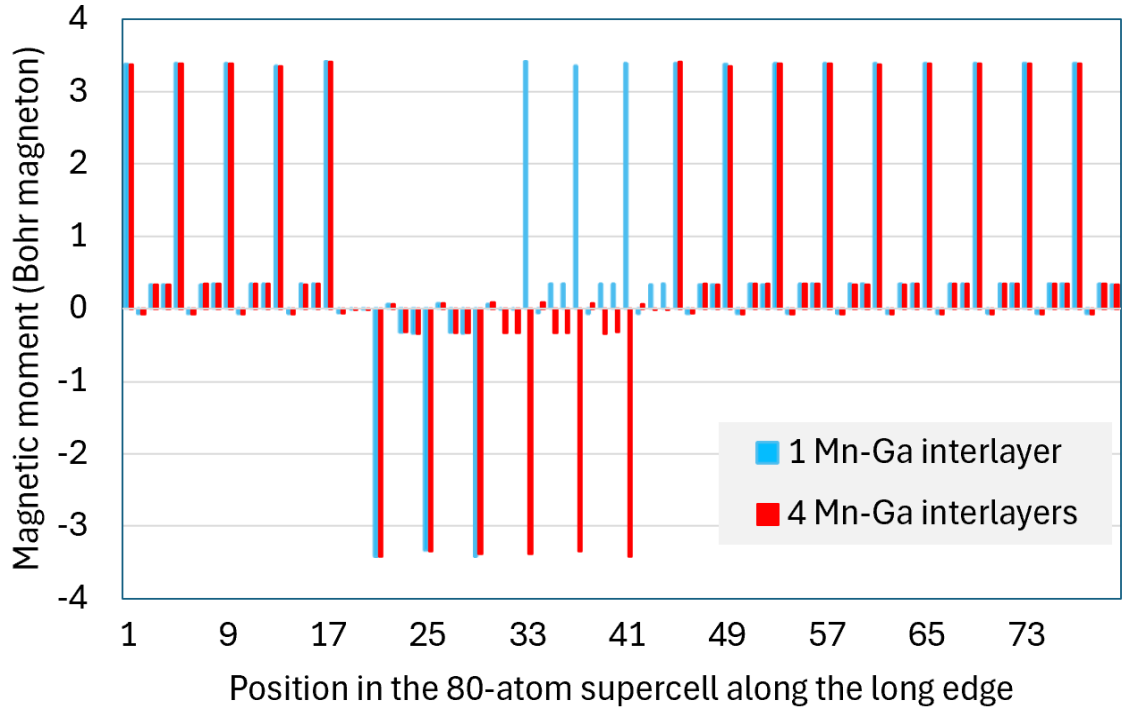

*Figure S1. The atomic magnetic moments vs position within a supercells with 80 atoms. The blue (red) bars indicate moments in a supercell with two APBs separated by one (four) Mn-Ga layers. All magenic moments are collinear. The narrow antiparallel region has three (six) Mn-Ga layers. The largest (smallest) moments are located at the Mn (Ga) sites. The data with one Mn-Ga interlayer (blue) corresponds to the supercell shown in Fig. 2a.*

In search for perpendicular-to-surface components of magnetization, $M_z$, we performed DFT simulations of the Ni-Mn-Ga surface using the QuantumATK [31] simulation package. The model system consisted of a periodic slab structure comprising 96 atoms within a triclinic unit cell defined by lattice constants of a = 0.41 nm, b = 2.32 nm, c = 2.23 nm. The individual slabs are separated by vacuum. The supercell contains two APBs as shown in Fig. S2. Spin-polarized calculations were performed with an explicitly defined noncollinear magnetic configuration, where Mn moments at the APBs were initialized with a positive (negative) vertical component, $m_z$, in the first (second) APB. This noncollinear supercell of Fig. S2b was compared with a collinear magnetic order with zero vertical component of magnetization shown in Fig. S2a. The electronic and magnetic structure was computed using the linear combination of atomic orbitals (LCAO) method with norm-conserving pseudopotentials sourced from the PseudoDojo library [32]. Element-specific localized basis sets were used for Mn, Ni, and Ga atoms. The exchange-correlation functional employed was the Perdew–Burke–Ernzerhof generalized gradient approximation [27]. Numerical accuracy was ensured with a high-density real-space mesh cutoff of 105 Hartree, and a Monkhorst-Pack k-point mesh derived from a reciprocal-space density of 0.4 nm$^{-1}$ in each direction. Total energy computations were used to compare the stability of the magnetic configurations of Fig. S2. The strong AFM exchange interaction does prefer the collinear order of Fig. S2 (a), in agreement with our VASP results, despite the presence of the surface and vacuum gap.

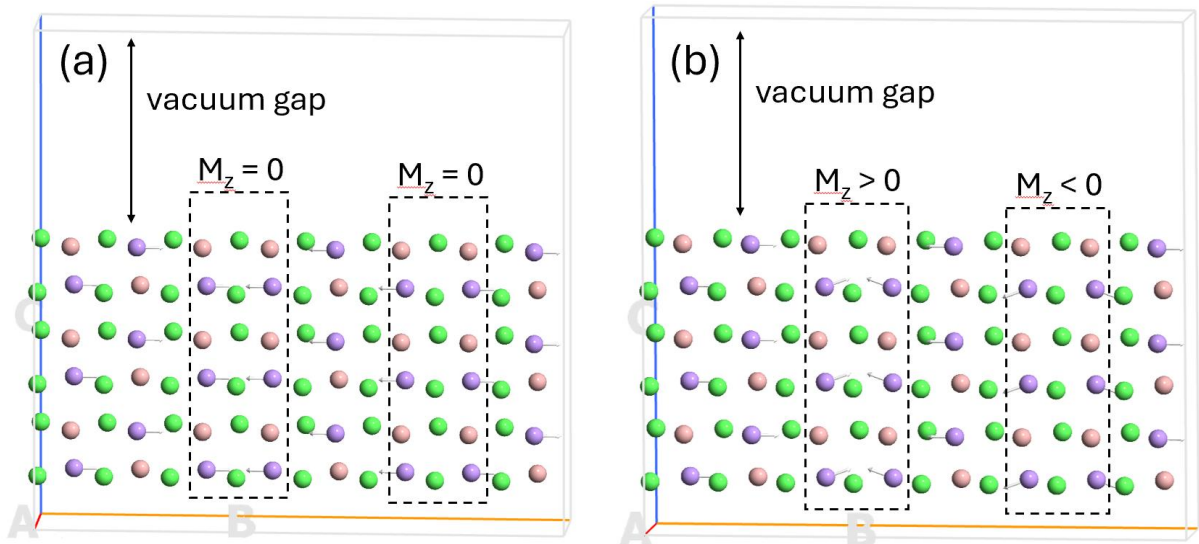

*Figure S2. Supercell comprising Ni$_2$MnGa slab and a vacuum gap used in Quantum ATK simulation of electronic energy for cases with collienar (a) and non-collinear (b) magentic moments. Both cases contain two planar APB in analogy to the VASP simulations. Only the case with one Mn-Ga interlayer between the two APBs is simulated. Mn atoms are in magenta, Ga in orange and Ni in green. The magentic moments are shown in gray.*

**Acknowledgment**

This work was supported by the Czech Science Foundation (project No. 24-11361S) and by the Ferroic Multifunctionalities project (FerrMion) [Project No. CZ.02.01.01/00/22008/0004591] by the Ministry of Education Youth and Sports (MEYS), co-funded by the European Union. Computational resources were provided by the e−INFRACZ project (ID: 90254) supported by MEYS.